%% file: ms.tex
\documentclass[preprint]{aastex}
\shorttitle{Exploiting Low-Dimensional Structure}
\shortauthors{Richards, Freeman, Lee, Schafer}
\usepackage{color}
\newcommand{\x}{{\bf x}}
\newcommand{\y}{{\bf y}}
\newcommand{\z}{{\bf z}}

\renewcommand{\P}{{\bf P}}

\begin{document}

\title{Exploiting Low-Dimensional Structure in Astronomical Spectra}
\author{Joseph W. Richards, Peter E. Freeman, Ann B. Lee, Chad M. Schafer}
\email{jwrichar@stat.cmu.edu}
\affil{Department of Statistics, Carnegie Mellon University, 5000 Forbes Avenue, Pittsburgh, PA 15213}

\begin{abstract}
Dimension-reduction techniques can greatly improve statistical inference in astronomy.
A standard approach is to use Principal Components Analysis (PCA).
In this work we apply a recently-developed technique, diffusion maps, to astronomical
spectra for data parameterization and dimensionality reduction, and
develop a robust, eigenmode-based framework
for regression.
We show how our framework provides a computationally efficient means by which
to predict redshifts of galaxies, and thus could
inform more expensive redshift estimators
such as template cross-correlation.  It also provides a natural means
by which to identify outliers (e.g., misclassified spectra, spectra
with anomalous features).
We analyze 3835 SDSS spectra and show how our framework
yields a more than 95\% reduction in dimensionality.
Finally, we show that the prediction error
of the diffusion map-based regression approach is markedly smaller than that of a similar 
approach based on PCA, clearly demonstrating the superiority of diffusion
maps over PCA for this regression task.
\end{abstract}


\section{Introduction}

\label{sect:intro}

Galaxy spectra are classic examples of high-dimensional data, with
thousands of measured fluxes providing 
information about the physical conditions of the observed object.
To make computationally efficient inferences about these 
conditions, we need to first reduce the dimensionality of the data 
space while preserving relevant physical information. 
We then need to find simple relationships between the reduced data and physical parameters of
interest.
Principal Components Analysis (PCA, or the Karhunen-Lo\`eve transform) is a standard method for the first step; its application to astronomical spectra is described in, e.g., \citet{BorosonGreen1992},
\citet{Connolly1995}, \citet{Ronen1999}, \citet{Folkes1999},
\citet{Madgwick2003}, \citet{Yip2004a},
\citet{Yip2004b}, \citet{Li2005}, \citet{Zhang2006}, 
\citet{VDB2006}, \citet{Rogers2007}, and \citet{ReFiorentin2007}.
In most cases, the authors do not proceed to the second step but only
 ascribe physical significance to the first few eigenfunctions from PCA
(such as the ``Eigenvector 1" of \citeauthor{BorosonGreen1992}).
Notable exceptions are \citeauthor{Li2005}, \citeauthor{Zhang2006},
and \citeauthor{ReFiorentin2007} However, 
as we discuss in {\S}\ref{sect:app}, these authors combine 
eigenfunctions in an ad hoc manner with no formal methods or
statistical criteria for regression and risk (i.e., error) estimation.

In this work we present a unified framework for regression and data parameterization of astronomical spectra. The main idea is to describe 
the important structure of a data set in terms of its 
{\em fundamental eigenmodes}.
The corresponding eigenfunctions are used both as coordinates for the data 
and as orthogonal basis functions for regression.  
We also introduce the {\em diffusion map} framework 
(see, e.g., \citealt{Coifman:Lafon:06}, \citealt{LafonLee2006}) 
to astronomy, comparing and contrasting it with PCA for regression analysis of SDSS galaxy spectra.  PCA is a global method that finds linear low-dimensional 
projections of the data; it attempts to preserve Euclidean distances between all data points and is often not robust to outliers. 
The diffusion map approach, on the other hand, is non-linear and instead retains distances that reflect the (local) connectivity of the data.  
This method is robust to outliers and is often able to unravel the intrinsic geometry and the natural (non-linear) coordinates of the data.

In {\S}\ref{sect:diff} we describe the diffusion map method for data
parameterization.
In {\S}\ref{sect:regress} we introduce the technique of {\em adaptive regression} using eigenmodes.
In {\S}\ref{sect:app} we demonstrate the effectiveness of our proposed PCA- and
diffusion-map-based regression techniques for 
predicting the redshifts of SDSS spectra.
Our PCA- and diffusion-map-based approaches provide a fast and
statistically rigorous means of identifying 
outliers in redshift data. The returned embeddings also provide an
informative visualization of the results.  In {\S}\ref{sect:summary} we summarize our results.

\section{Diffusion Maps and Data Parameterization}

\label{sect:diff}
\begin{figure}
\epsscale{0.7}
\plotone{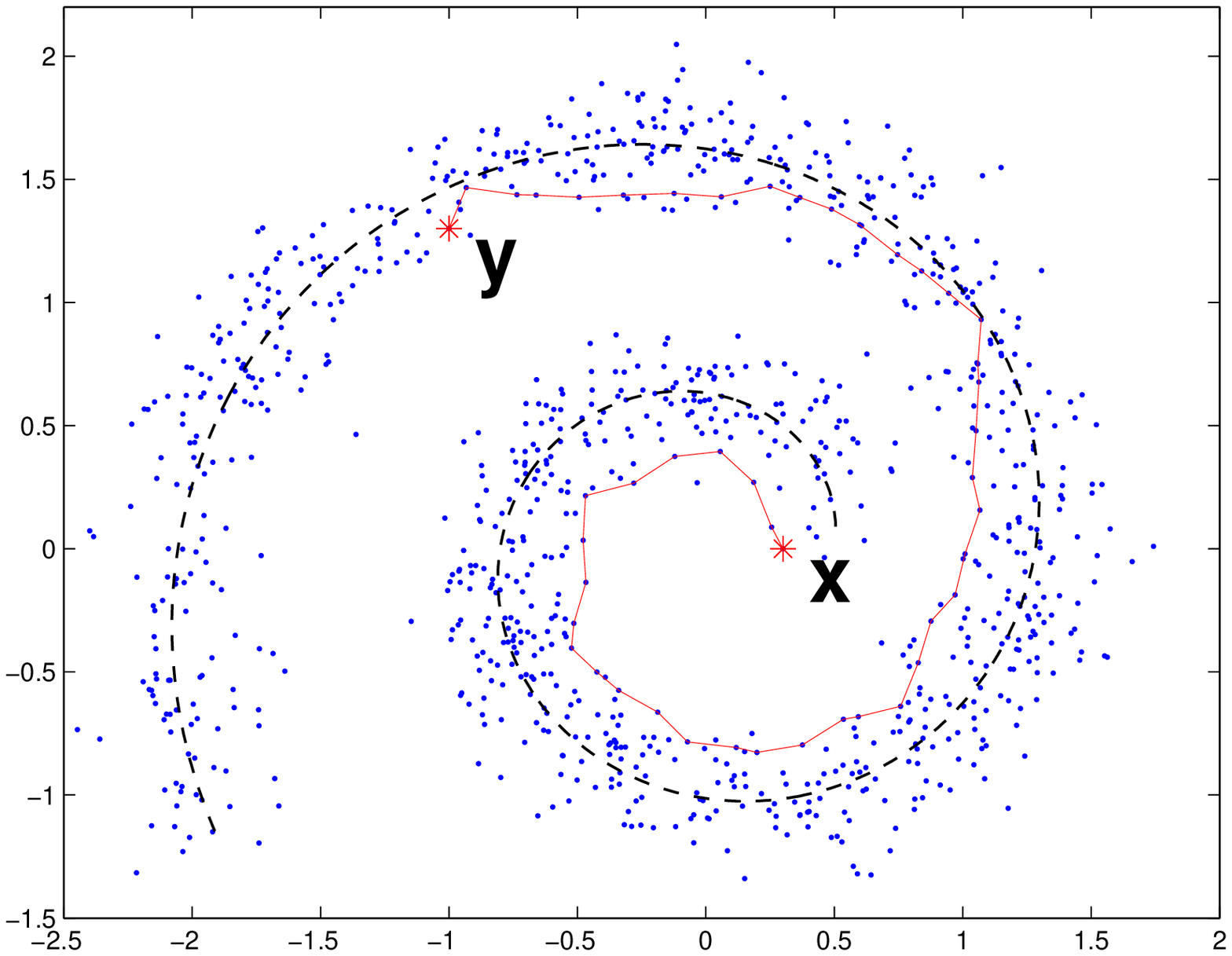}
\vspace{0.7in}
\epsscale{0.9}
\plottwo{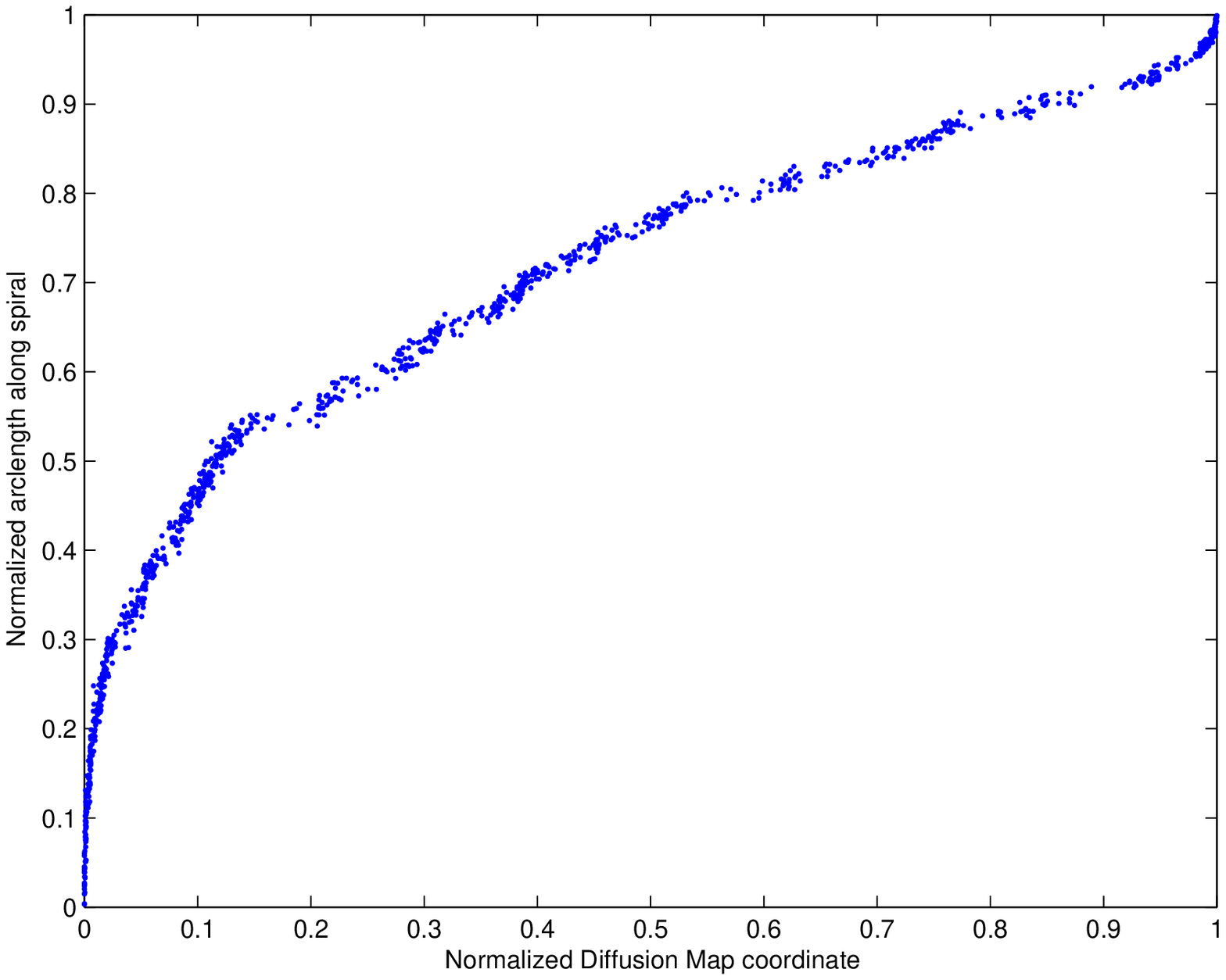}{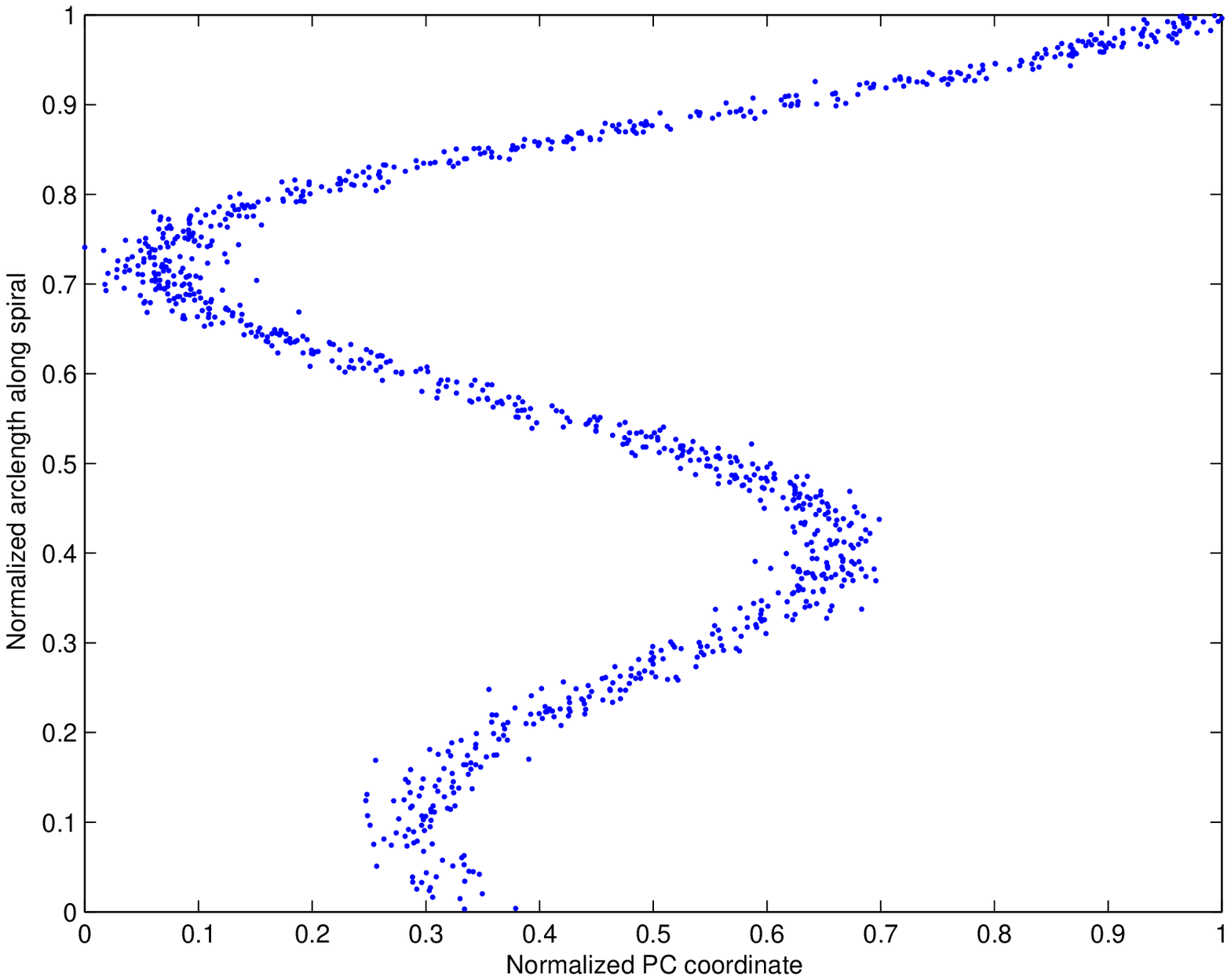}
\caption{An example of a one-dimensional manifold (dashed line) with Gaussian noise embedded in
two or higher dimensions.  The path (solid line) from $\x$ to $\y$ reflects the natural geometry of
the data set which is captured by the
diffusion distance between $\x$ and $\y$. 
The plot on the lower left shows that the first diffusion map coordinate is a monotonically increasing
function of the
arc length of the spiral; this is not the case in the
lower right plot, which shows the same relationship for the first PC coordinate.}
\label{fig:spiral}
\end{figure}

The variations in a physical system can sometimes be described by
a few parameters, while measurements of the system are
necessarily of very high dimension; geometrically, the data are
points in the $p$-dimensional space $\mathbb{R}^p$, with $p$ large.
In our case, a data point is a galaxy spectrum, with the 
dimension $p$ given by the number of wavelength bins ($p \gtrsim 10^3$),
and a full data set could consist of hundreds of thousands of spectra.
To make inference and predictions tractable, 
one seeks to find a simpler parameterization of the system. The most 
common method for dimension reduction and data parameterization 
is Principal Component Analysis (PCA), where the data are projected 
onto a lower-dimensional hyperplane. For complex situations, 
however, the assumption of linearity may lead to sub-optimal 
predictions. A linear model pays very little attention to the 
natural geometry and variations of the system. The top plot in Figure
\ref{fig:spiral} 
illustrates this clearly by showing a data 
set that forms a one-dimensional noisy spiral in $\mathbb{R}^2$. 
Ideally, we would like to find a coordinate system that reflects 
variations along the spiral direction, which is indicated by the
dashed line. It is obvious that any
projection of the data onto a line would be unsatisfactory.  Results
of a PCA analysis of the noisy spiral are shown in the lower-left plot
in Figure \ref{fig:spiral}.

In this section, we will use diffusion maps 
(\citeauthor{Coifman:Lafon:06}, \citeauthor{LafonLee2006}) --- a non-linear technique --- 
to find a natural coordinate system for the data. 
When searching for a lower-dimensional description, one needs to decide 
what features to preserve and what aspects of the data one is 
willing to lose. The diffusion map framework attempts to retain 
the cumulative local interactions between its data points, or 
their ``connectivity" in the context of a fictive diffusion process over the data. 
We demonstrate how this can be a better method to learn 
the intrinsic geometry of a data set than by using, e.g., PCA. 

Our strategy is to first define a distance metric $D(\x,\y)$ that reflects
the connectivity of two points $\x$ and $\y$, then find a map to a 
lower-dimensional space (i.e., a new data parameterization) that 
best preserves these distances. 
(As before, a ``point'' in $p$-dimensional space represents
a complete astronomical spectrum of $p$ wavelength bins.)
The general idea is that we call two data points ``close'' if there 
are many short paths between $\x$ and $\y$ in a jump diffusion process between data points.
In Figure \ref{fig:spiral}, the Euclidean distance
between two points is an inappropriate measure of
similarity. If, instead, one imagines a random walk starting at ``$\x$,'' and
only stepping to immediately adjacent points, it is clear that 
the time it would take for that walk to reach ``$\y$''  would reflect
the length along the spiral direction.  This latter distance measure
is represented by the solid path from $\x$ to $\y$ in Figure \ref{fig:spiral}.
We will make this measure of connectivity formal in what follows.

The starting point is to construct a weighted graph where the
nodes are the observed data points. 
The weight given to the edge connecting $\x$ and $\y$ is
\begin{equation}
w(\x,\y) = \exp\left(-\frac{s(\x,\y)^2}{\epsilon}\right),
\label{eqn:diffw}
\end{equation}
where $s(\x,\y)$ is a locally relevant similarity measure.
For instance, $s(\x,\y)$ could be chosen as 
the Euclidean distance between $\x$ and $\y$ (denoted here $\|\x-\y\|$)
when $\x$ and $\y$ are vectors.
But, the choice of $s(\x,\y)$ is not crucial, and this gets to the heart
of the appeal of this approach:
it is often simple to determine whether or not two data points are 
``similar'', 
and many choices of $s(\x,\y)$ will suffice for measuring this
local similarity.
The tuning parameter $\epsilon$ is chosen small enough that
$w(\x,\y) \approx 0$ unless $\x$ and $\y$ are similar,
but large enough such that the constructed graph is fully connected.

The next step is to use these weights to build a Markov random walk on 
the graph. From node (data point) $\x$, the probability of stepping
directly to $\y$ is defined naturally as 
\begin{equation}
p_1(\x,\y) = \frac{w(\x,\y)}{\sum_{\z}w(\x,\z)}.
\label{eqn:diffp}
\end{equation}
This probability is close to zero unless $\x$ and $\y$ are similar. Hence, in
one step the random walk will move only to very similar nodes (with high
probability). These one-step transition probabilities are stored in the $n$ by $n$ 
matrix $\P$.
It follows from standard theory of Markov chains (\citealt{KemenySnell1983}) that, for a positive integer $t$, the element 
$p_t(\x,\y)$ of
the matrix power $\P^t$ gives the probability of
moving from $\x$ to $\y$ in $t$ steps.
Increasing $t$ moves the random walk
forward in time, propagating the local influence of a data point 
(as defined by the kernel $w$)
with its neighbors.

For a fixed time (or scale) $t$, $p_t(\x,\cdot)$ is a vector representing
the distribution after $t$ steps of the random walk over the nodes of the
graph, conditional on the
walk starting at $\x$.
In what follows, the points $\x$ and $\y$ are
close if the conditional distributions 
$p_t(\x,\cdot)$ and $p_t(\y,\cdot)$, are similar. 
Formally, the diffusion distance at a scale $t$ is defined as
\begin{equation}
D_t^2(\x,\y) = \sum_{\z} \frac{\left(p_t(\x,\z) - p_t(\y,\z)\right)^2}{\phi_0(\z)}
\label{eqn:diffdist}
\end{equation}
where $\phi_0(\cdot)$ is the stationary distribution of the random walk, i.e.,
the long-run proportion of the time the walk spends at
node $\z$. 
Dividing by $\phi_0(\z)$ serves to reduce the influence of nodes
which are visited with high probability regardless of the starting point of the
walk.
The distance $D_t(\x,\y)$ will be small only if $\x$ and $\y$ are connected by
many short paths with large weights.  This construction of
a distance measure is robust to noise and outliers because it
simultaneously accounts for the cumulative effect of {\em all} paths between the
data points. 
Note that the geodesic distance (the shortest path in a graph), on the other hand, often takes shortcuts due to noise.


The final step is to find a low-dimensional embedding of the data where Euclidean distances reflect diffusion distances.
A biorthogonal spectral decomposition of the matrix $\P^t$ gives
$p_t(\x,\y) = \sum_{j \ge 0} \lambda_j^t \psi_j(\x) \phi_j(\y)$,  
where $\phi_j$, $\psi_j$, and $\lambda_j$, respectively, represent left eigenvectors, right eigenvectors and eigenvalues 
of $\P$. It follows that
\begin{equation}
D^2_t(\x,\y)~= ~\sum_{j=1}^{\infty} \lambda_j^{2t}(\psi_j(\x)-\psi_j(\y))^2.\label{eq:Dt}
\end{equation}
 The proof of Equation~\ref{eq:Dt} and the details of the computation
 and normalization of the eigenvectors  $\phi_j$ and $\psi_j$ are given in
 \citeauthor{Coifman:Lafon:06} and
 \citeauthor{LafonLee2006}.\footnote{Sample code in Matlab and R for
   diffusion maps at {\tt  http://www.stat.cmu.edu/\~{}annlee/software.htm}}  By retaining the
 $m$ eigenmodes corresponding to the $m$ largest nontrivial
 eigenvalues and by introducing the diffusion map
\begin{equation}
\Psi_t: \x \mapsto [\lambda_1^t\psi_1(\x), \lambda_2^t\psi_2(\x), \cdots,\lambda_m^t\psi_m(\x)]
\label{eqn:diffusion_map}
\end{equation}
from $\mathbb{R}^p$ to $\mathbb{R}^m$, we have that 
\begin{equation}
D^2_t(\x,\y)~\simeq ~\sum_{j=1}^m \lambda_j^{2t}(\psi_j(\x)-\psi_j(\y))^2 ~=~||\Psi_t(\x) - \Psi_t(\y)||^2 \,,
\label{eqn:diffpres}
\end{equation}
i.e., Euclidean distance in the $m$-dimensional embedding defined by equation~\ref{eqn:diffusion_map}
approximates diffusion distance.
In contrast, Euclidean distances in PC maps approximate the original
Euclidean distances $\|\x-\y\|$.
Again, consider the example in Figure \ref{fig:spiral}.
The plot on the lower left shows that the first diffusion map coordinate is a monotonically increasing
function of the
arc length of the spiral; this is not the case in the
lower right plot, which shows the same relationship for the first PC coordinate. Indeed, the relationship
with the first PC coordinate is not even one-to-one.

The choice of the parameters $m$ and $t$ is determined by the fall-off of the eigenvalue spectrum as well 
as the problem at hand (e.g., clustering, classification, regression, 
or data visualization).  An objective measure
of performance should be defined and utilized to find data-driven best choices for these tuning parameters. 
In this work, the final goal
is regression and prediction of redshift. In the next section, we show how the number of coordinates, $m$, can 
be chosen by cross-validation, once one has defined an appropriate statistical ``risk" function. The particular 
choice of $t$, on the other hand, will not matter in the regression framework, as it will only represent a 
rescaling of the $m$ selected basis vectors.

\section{Adaptive Regression Using Orthogonal Eigenfunctions}
\label{sect:regress}
Our next problem is how to, in a statistically rigorous way, predict a function $y=r(\mathbf{x})$ (e.g., redshift, age, or metallicity of galaxies) of data 
(e.g., spectrum $\mathbf{x}$) in very high dimensions using a sample
of known pairs ($\x,y$). As before, imagine that our data are points in $\mathbb{R}^p$, but that the
natural variations in the system are along a low dimensional space $\mathcal{X} \subset \mathbb{R}^p$.
The set $\mathcal{X}$ could, for example, be a non-linear submanifold embedded in $\mathbb{R}^p$.
In our toy example in Figure \ref{fig:spiral}, $\mathcal{X}$ is the one-dimensional spiral, but the data are observed
in $p=2$ dimensions.
The key idea is that one may view the eigenfunctions from PCA or diffusion maps
(a) as {\em coordinates} of the data points, as shown in the previous section,
or (b) as forming a {\em Hilbert orthonormal basis} for any function (including the regression function $r(\mathbf{x})$) supported on the 
subset $\mathcal{X}$. Rather than applying an arbitrarily chosen prediction scheme in the computed diffusion or PC space (as in, e.g., \citeauthor{Li2005}, \citeauthor{Zhang2006}, and \citeauthor{ReFiorentin2007}), we utilize the latter insight to formulate a general regression and risk estimation framework. 

Any function $r$ satisfying $\int r(\x)^2 dx < \infty$, where $\x \in \mathcal{X} $, can be written as
\begin{equation}
r(\x) = \sum_{j=1}^{\infty} \beta_j \psi_j(\x) \,, 
\label{eqn:orthonorm}
\end{equation}
where the sequence of functions $\{\psi_1,\psi_2,\cdots\}$ forms an
orthonormal basis.  The choice of basis functions is traditionally {\em not} adapted to the geometry of the data, or the set $\mathcal{X}$.
Standard choices are, for example, Fourier or wavelet bases for $\mathbf{L}^2(\mathbb{R}^p)$, which are constructed as tensor 
products of one-dimensional bases. The latter approach makes sense for low dimensions, for example for $p=2$, but quickly becomes
intractable as $p$ increases (see, e.g., \citealt{Bellman:61} for the ``curse of dimensionality''). In particular, note that if a wavelet basis
in one dimension consists of $q$ basis functions, and hence 
requires the estimation of $q$ parameters, the naive tensor basis in $p$ dimensions will have $q^p$ basis functions/parameters,
creating an impossible inference problem even for moderate $p$.
Because this basis is not adapted to $\mathcal{X}$, there is little hope of 
finding a subset of these basis functions which will
do an adequate job of modeling the response.

In this work, we propose a new adaptive framework where the basis functions reflect the intrinsic geometry of the data.  Furthermore, we use a formal statistical method to estimate the risk and the optimal parameters in the model. First, rather than using a generic tensor-product basis for the high-dimensional space $\mathbb{R}^p$, we 
construct a data-driven 
basis for the lower-dimensional, possibly non-linear set $\mathcal{X}$ where the data lie. 
Let $\{{\psi_1},{\psi_2},\cdots,{\psi_n}\}$ be the orthogonal eigenfunctions computed by PCA or diffusion maps. 
Our regression function estimate $\widehat{r}(\x)$ is then given by 
\begin{equation}
\widehat{r}(\x) = \sum_{j=1}^{m} \widehat{\beta_j} {\psi_j}(\x), 
\label{eqn:orthoreg}
\end{equation}
where the different terms in the series expansion represent the
fundamental eigenmodes of the data, and $m \leq n$ is chosen to
minimize the prediction risk that we will now define rigorously.

\subsection{Risk: Theory and Estimation}
\label{sect:risk}

A key aspect of our approach is that the choice of the models is driven by the minimization of a well-justified, objective error criterion
which compensates for overfitting. This is critical, as any basis could be utilized to fit the observed data well; this does not provide,
however, any assurance that the model applies beyond these data.
To begin, we establish the standard stochastic framework within which regression models are assessed.
We are given $n$ pairs of observations $(X_1,Y_1), \ldots, (X_n, Y_n)$, with the task of predicting the 
response $Y=r(X)+\epsilon$ at a new data point $X=\x$, where $\epsilon$ represents random noise.  
(In {\S}\ref{sect:app}, the response $Y$ is the redshift, $z$, and $X$ is a complete spectrum.) 
In nonparametric regression by orthogonal functions, 
one assumes that $r(\x)$ is given 
according to equation~(\ref{eqn:orthonorm}), with its estimator given
by equation~(\ref{eqn:orthoreg}), with $m \leq n$ where $\{\psi_j\}$
is a fixed basis.
The primary goal is to minimize the
{\em prediction risk} (i.e., expected error), commonly quantified by
the mean-squared error (MSE)
\begin{equation}
R(m)=\mathbb{E}[Y-\widehat{r}(X)]^2,
\label{eqn:MSE}
\end{equation} 
where the average is taken over all possible realizations of $(X,Y)$,
including the randomness in the evaluation points $X$, the
responses $Y$, and the estimates $\widehat{\beta_j}$.
Thus, $\mathbb{E}[\cdot]$ averages everything that is random, including the randomness in the evaluation points $X$
and the randomness in the estimates $\widehat{\beta_j}$. This leads to protection against overfitting: if a basis
function $\psi_j$ is unnecessarily included in the model, 
its coefficient $\widehat{\beta_j}$ will only add variability 
or variance to
$\widehat{r}(X)$ and not improve the fit, hence increasing $R(m)$.
(On the other hand, as $m$ becomes too small, 
the estimator becomes increasingly biased, also increasing $R(m)$.)
Thus, the ideal choice of $m$ is neither too large, nor too small.
In nonparametric statistics, this is dubbed the ``bias-variance tradeoff"
(see, e.g., \citealt{Wasserman2006}).
A secondary goal is {\em sparsity}; more specifically, 
among the estimators with a small risk, 
we prefer representations with a smaller $m$.

Since $R(m)$ is a population quantity, one needs to appropriately estimate it from the data. 
An estimate based on the full data set will underestimate the error and lead to a model with high bias. 
Here we will use the method of $K$-fold cross-validation 
(see, e.g., \citeauthor{Wasserman2006}) to achieve 
a better estimate of the prediction risk. The basic idea is to randomly split the data set into $K$ blocks 
 of approximately the same size; $K=10$ is a common choice. For $k=1$ to $K$, we delete block $k$ from the data. We then fit the model to the 
remaining $K-1$ blocks and compute the observed squared error $\widehat{R}_{(-k)}(m)$ on the $k$th block which was not included in the fit. The CV estimate of the risk is defined as $\widehat{R}_{CV}(m)=\frac{1}{K}\sum_{k=1}^{K} \widehat{R}_{(-k)}(m)$.
It can be shown that this quantity is an approximately unbiased estimate of the true error $R(m)$.
Thus, we choose the model parameters that minimize the CV estimate $\widehat{R}_{CV}(m)$ of the risk, i.e., 
we take $m_{\rm opt} = \arg \min \widehat{R}_{CV}(m)$.

Finally, we note that the ideas of CV introduced here generalize to cases where the model
parameters are of higher dimension. For example, in the diffusion
map case, the risk is minimized over both the bandwidth $\epsilon$ and the number of eigenfunctions $m$. The CV estimate of the
risk is implemented in the same fashion, but the search space for finding the minimum is larger.
In what follows, the notation will make it clear which
model parameters we are minimizing over by writing, for
example, $R(\epsilon, m)$.

To summarize, our claim is that the proposed regression framework will lead to efficient inference in high 
dimensions, as we are effectively performing regression in a lower-dimensional space $\mathcal{X}$ that 
captures the natural variations of the data, where the optimal
dimensionality is chosen to minimize prediction risk in our regression
task. Finally, the use of eigenfunctions in both the data parameterization 
and in the regression formulation provides an elegant, unifying framework for analysis and prediction. 
 

\section{Redshift Prediction Using SDSS Spectra}

\label{sect:app}

We apply the formalism presented in {\S}{\S}\ref{sect:diff}-\ref{sect:regress}
to the problem of predicting redshifts for a sample of SDSS spectra.
Physically similar objects residing at similar redshifts will have
similar continuum shapes as well as absorption lines occurring at
similar wavelengths.  Hence the 
Euclidean distances between their spectra will be small.
 The proposed regression framework with diffusion map or PC
 coordinates provides a natural means by which to predict
redshifts.  Furthermore, it is computationally efficient, making its
use appropriate for large databases such as the SDSS;
one can use these predictions to
inform more computationally expensive techniques by narrowing down
the relevant parameter space (e.g., the redshift range or the 
set of templates in cross-correlation techniques).  
Adaptive regression also provides a useful 
tool for quickly identifying anomalous data points (e.g., objects
misclassified as galaxies), galaxies that have relatively rare
features of interest, and 
galaxies whose SDSS redshift estimates may be incorrect.
 
\subsection{Data Preparation}

Our initial data sample consists of spectra that are classified as galaxies
from ten arbitrarily chosen spectroscopic plates of SDSS DR6
(0266$-$0274 inclusive, and 0286; \citealt{Adelman2008}).  
We remove spectra from this sample by applying three cuts.  The first
is motivated by aperture considerations: we analyze only those spectra
with SDSS redshift estimates $z_{\rm SDSS} \geq$ 0.05.  
To include spectra  with $z_{\rm SDSS} < 0.05$
would be to add an extra source of variation that would
adversely impact regression analysis.  The second cut is based on bin flags.
To avoid calibration issues observed at both the low and high
wavelength ends, we remove the first 100 and last 250 wavelength bins
from each spectrum;
then we determine what proportion of the remaining 3500 bins are flagged
as bad.  If this proportion exceeds 10\%, we remove the spectrum from the
sample; if not, we retain the reduced spectrum for further analysis.  
We provide details on the third cut below.
The application of these cuts reduces our sample size from 5057
to 3835 galaxies.


We further process each spectrum in our sample as follows.
\begin{itemize}
\item We replace the flux values in the vicinity of
prominent atmospheric lines at 5577~\AA, 6300~\AA, and 6363~\AA~with
the sample mean of the nine closest bins on either side of each line.
The flux errors are estimated by averaging (in quadrature)
the standard errors of the fluxes for these bins.
\item We similarly replace the flux values in each bin flagged by SDSS as
part of an emission line, with flux and flux error estimates based
upon the closest 50 bins on either side of the line.  (Within this group
of 100 bins, we do not include those that are themselves flagged as
emission lines.)
We do this because highly variable emission line strengths 
can strongly bias distance calculations.
\item Last, after replacing flux values as necessary, we normalize
each spectrum to sum to 1 to mitigate variation due to differences in luminosity between 
similar galaxies at similar redshifts.
\end{itemize}

\begin{figure}
\epsscale{0.75}
\plotone{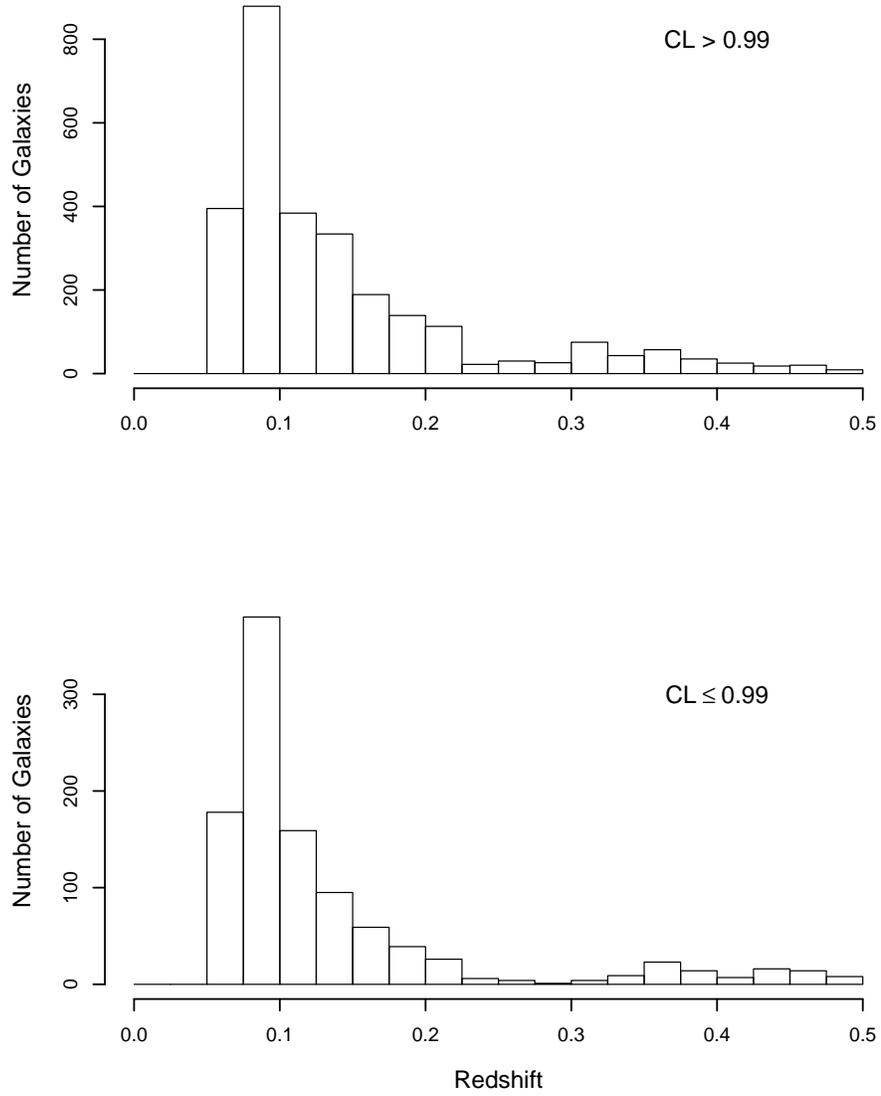}
\caption{Distributions of SDSS redshift estimates in our
high-CL (top) and low-CL (bottom) samples.  We train our regression 
model using the 2793 high-CL galaxies only, then apply those
predictions to the 1042 low-CL galaxies.}
\label{fig:zdesign}
\end{figure}

In its data reduction pipeline, SDSS estimates spectroscopic redshifts,
$z_{\rm SDSS}$, standard errors, $\sigma_{z_{\rm SDSS}}$, and
``confidence levels," CL, the latter of which are functions
of the strengths of observed lines (and thus should
not be interpreted probabilistically).\footnote{
See {\tt http://www.sdss.org/dr6/algorithms/redshift\_type.html}.}
Lacking knowledge of the true redshifts in our sample, we use 
$z_{\rm SDSS}$ and $\sigma_{z_{\rm SDSS}}$ to fit our regression model.
Since poorly estimated redshifts can bias the model,
we divide our data sample into two groups, fitting with only
those 2793 galaxies with CL $>$ 0.99.
We then use the fitted model to predict redshifts for the other 1042 galaxies.
(It is here that we make our third data cut: to avoid issues of extrapolation,
we removed 19 of 1061 spectra with CL $\leq$ 0.99 whose SDSS redshift estimates
lie outside the range of our training set, i.e. those with $z_{\rm
  SDSS} > 0.50$.)  As shown in Figure \ref{fig:zdesign}, the distributions of
redshifts in our high- and low-CL samples are similar, implying that 
predicted redshifts for low-CL galaxies from the model built on
high-CL galaxies should not be systematically biased.

\subsection{Analysis}
\label{sect:anal}



In this section, we perform both PCA and diffusion map for our sample
and predict redshift using the
regression model introduced in {\S}\ref{sect:regress}.  We provide
details on the PCA algorithm in Appendix \ref{sect:pca}.

In the diffusion map analysis, 
we begin by calculating Euclidean distances between spectra
\begin{equation}
   s(\x, \y)~=~ \sqrt{\sum_k (f_{\x,k}-f_{\y,k})^2} \,,
\end{equation}
where $f_{\x,k}$ and $f_{\y,k}$ are the normalized fluxes in bin $k$ of
spectra $\x$ and $\y$, respectively.  We use these distances and a
chosen value of
$\epsilon$ to construct both the weights for the graph (see equation
\ref{eqn:diffw}) and the transition 
matrix $\P$ (see equation \ref{eqn:diffp}), from which eigenmodes are
generated.  Below we
discuss how we select the optimal value of $\epsilon$.  
As stated in {\S}\ref{sect:diff}, the value of the parameter $t$ 
(see equation \ref{eqn:diffusion_map}) is unimportant in
the context of regression, as any change in $t$ would be met 
with a corresponding
rescaling of the coefficients $\widehat \beta_j$ in the regression model,
such that predictions are unchanged.

\begin{figure}
\epsscale{1}
\plottwo{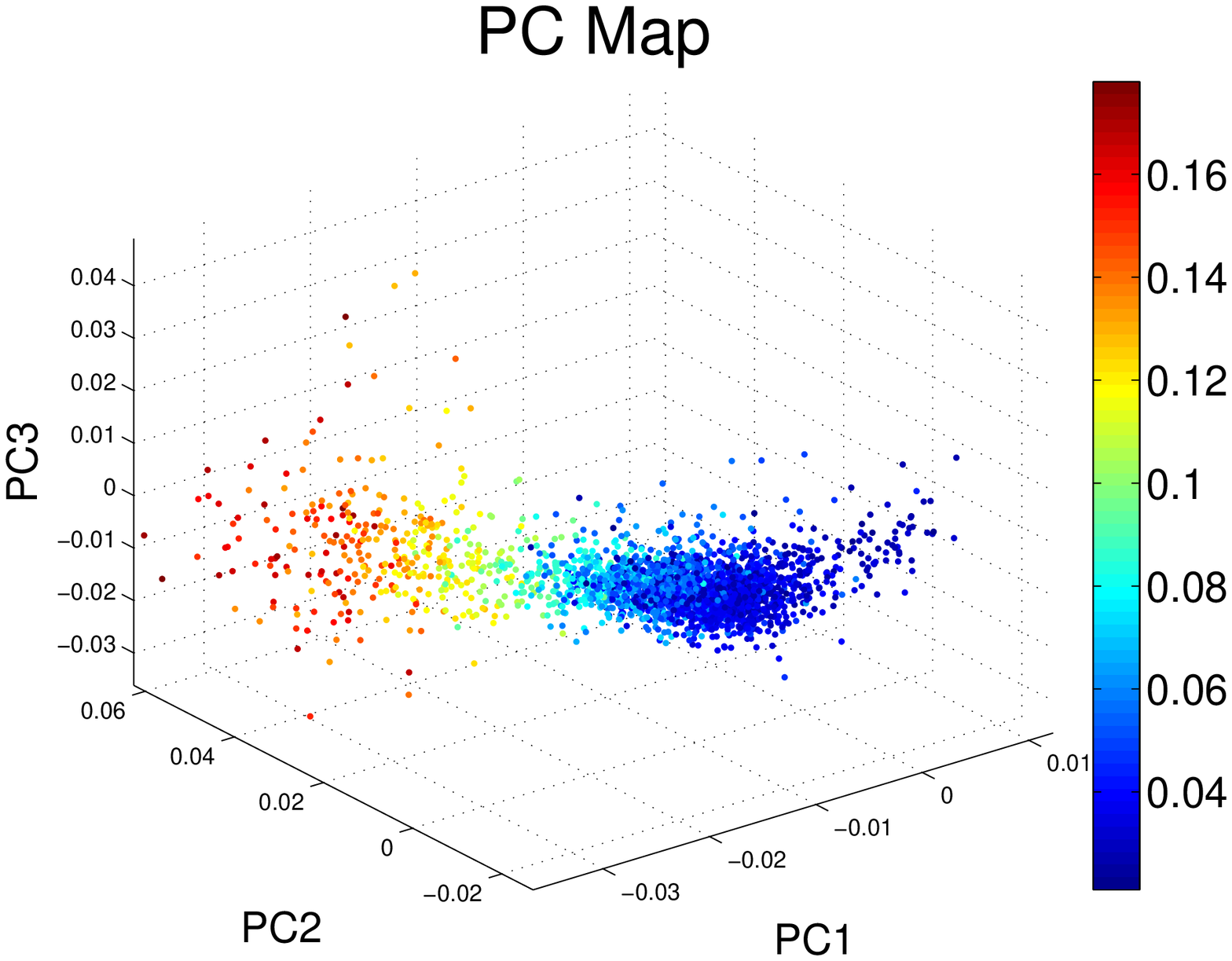}{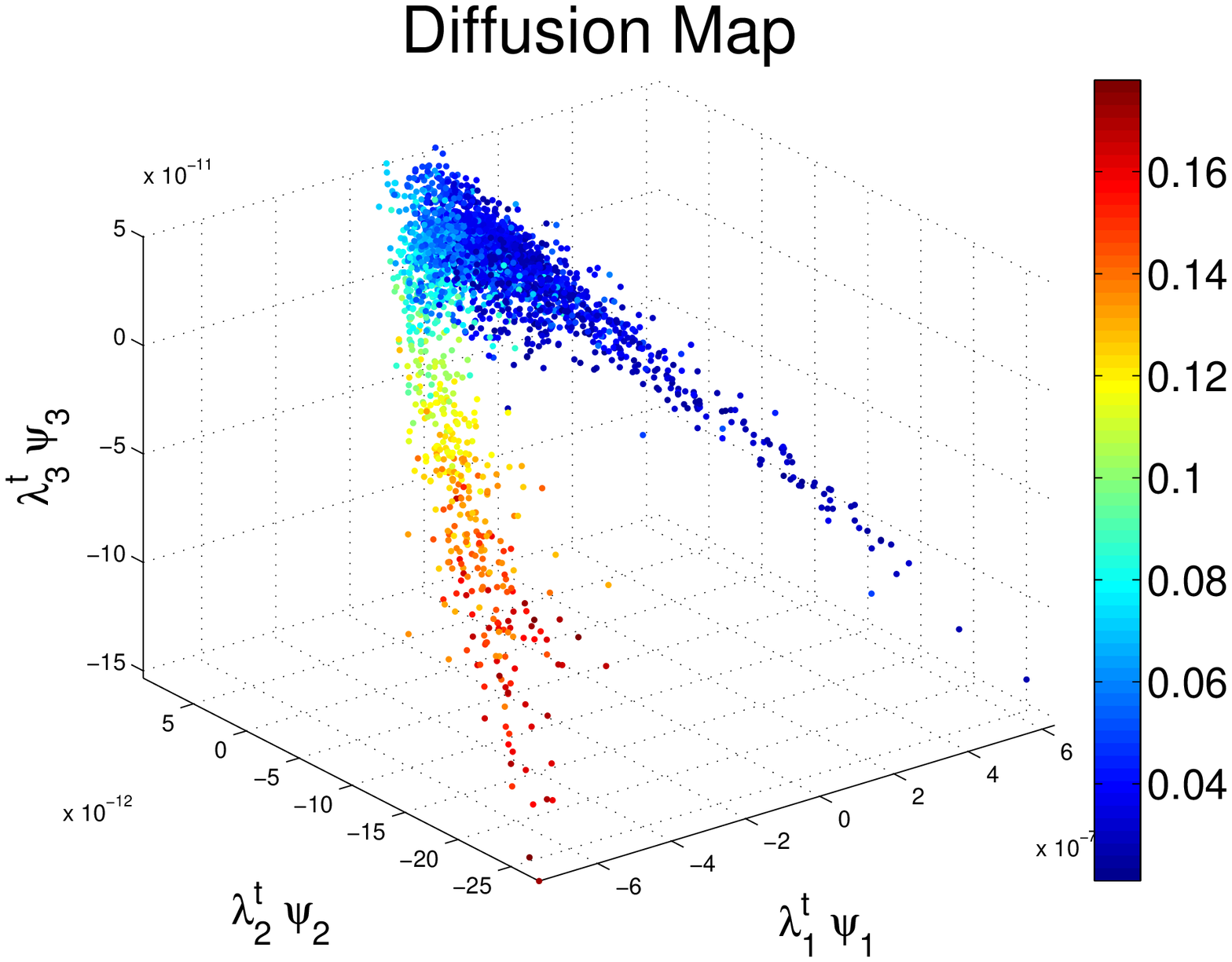}
\caption{Embedding of our sample of 2793 SDSS galaxy spectra with
  SDSS $z$ CL $> 0.99$ with
the first 3 PC and the first 3 diffusion map coordinates, respectively.
The color codes for $\log_{10}(1+z_{\rm SDSS})$ values.  Both
maps show a clear correspondence with redshift.}
\label{fig:zmaps}
\end{figure}

In Figure \ref{fig:zmaps} we plot the embedding of
the 2793 galaxies with CL $>$ 0.99
in the first three PC and diffusion map 
coordinates (e.g., $\lambda_i^t\psi_i(\cdot)$ in equation \ref{eqn:diffusion_map}).
We observe that the structure of each of these reparameterizations of
the original data corresponds in a simple way to $\log_{10}(1+z_{\rm
  SDSS})$.  These embeddings are a useful way to visualize the data
and to qualitatively identify subgroups of data and peculiar data points.

\begin{figure}
\epsscale{0.75}
\plotone{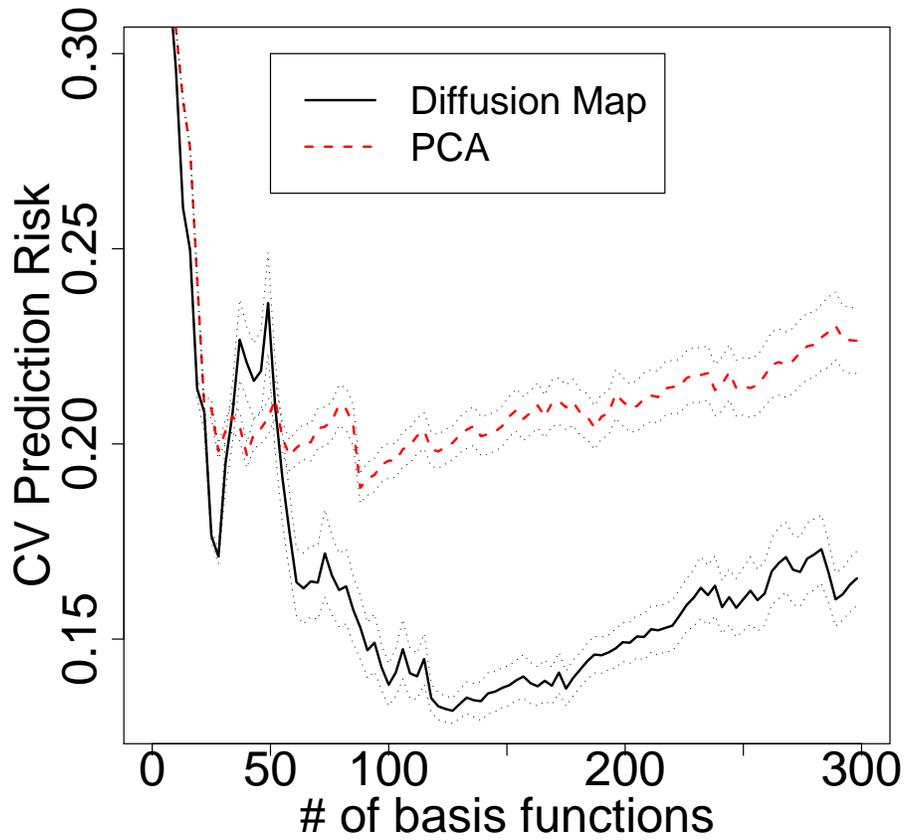}
\caption{Risk estimates ($\widehat{R}_{CV}$) for regression of $z$ on diffusion 
  map coordinates and PCs. Diffusion map attains a lower 
  risk for almost every number of coordinates in the regression. It also 
  achieves a lower minimum risk as indicated by Table~\ref{tab:zreg}.
Risk estimates are based on 50 repetitions of 10-fold CV.  Thick lines
represent mean risk at that model size and thin dotted lines are +/- 1
standard deviation bands.}
\label{fig:zrisk}
\end{figure}

In the next stage of analysis we use the computed eigenfunctions to
predict $z$ for our sample of 3835 galaxies.
We regress $z_{\rm SDSS}$ upon the diffusion map (and PC) eigenmodes
(cf.~equation~\ref{eqn:orthoreg}, where $\widehat r$ represents
our redshift estimates), weighting each data point by the 
inverse variance of its $z_{\rm SDSS}$, 1/$\sigma_{z_{\rm SDSS}}^2$,
to account for the uncertainties in $z_{\rm SDSS}$ measurements.
We repeat this step for a sequence of 
$m$ (and $\epsilon$) values, determining the optimal values of each
by minimizing the prediction risk $R(\epsilon,m)$, 
estimated via ten-fold cross-validation (see equation~\ref{eqn:MSE}
and subsequent discussion).  It is in this regression step that
we clearly observe the advantage of using diffusion maps over
principal components.  In Figure \ref{fig:zrisk} we show that
diffusion map achieves significantly lower
CV prediction risk for most choices of model size $m$ and
obtains a much lower minimum $\widehat{R}_{\rm CV}$, i.e.,
the optimal low-dimensional diffusion map
representation of our data captures the trend in $z$ better than the
PC representation.  Note that the trend in $\widehat{R}_{\rm CV}$ for both
PC and diffusion map basis functions is to decrease with increasing
model size for small models and to increase with increasing model size
for larger models.  This is the ``bias-variance tradeoff" that was
referred to in {\S}\ref{sect:risk}: as the size (complexity) of our model
increases, the bias of the model decreases while the variance of the
model increases.  Prediction risk is the sum of the squared bias and
variance of a model, explaining the behavior observed in Figure
\ref{fig:zrisk}: for small models, increasing model size leads to
decrease in bias that overwhelms
increase in variance while for large models, increase in model size
produces minimal decrease in bias and relatively large increase in variance.

In Table \ref{tab:zreg}, we show the parameters for the
optimal (minimal $\widehat{R}_{\rm CV}$) diffusion map and PC regression models.
Note that since our original data were in 3500
dimensions, our optimal diffusion map model achieves 
a 96.4\% reduction in dimensionality.  If we were to choose
an arbitrary small model size as is often done in the literature, our
prediction risk estimates would be terrible.  For example, for model
sizes $m = 10$ and 20, the CV prediction risks for regression on PC
basis functions are 0.305 and 0.209, respectively (compared to optimal
value 0.193), while regression on
diffusion map basis functions yields $\widehat{R}_{\rm CV}$ of 0.295
and 0.191, respectively (compared to optimal value 0.134).  The choice of
$\epsilon$ in the diffusion map model also has a significant impact on
results.  For values of $\epsilon$ that are too small, CV risks are
extremely large because the data points are no longer connected in the
diffusion process and consequently large outliers occur in the
diffusion map parameterization.  Likewise, large values of $\epsilon$
yield large prediction risks due to the large weights given to
connections between dissimilar data points.

\begin{figure}
\epsscale{0.6}
\plotone{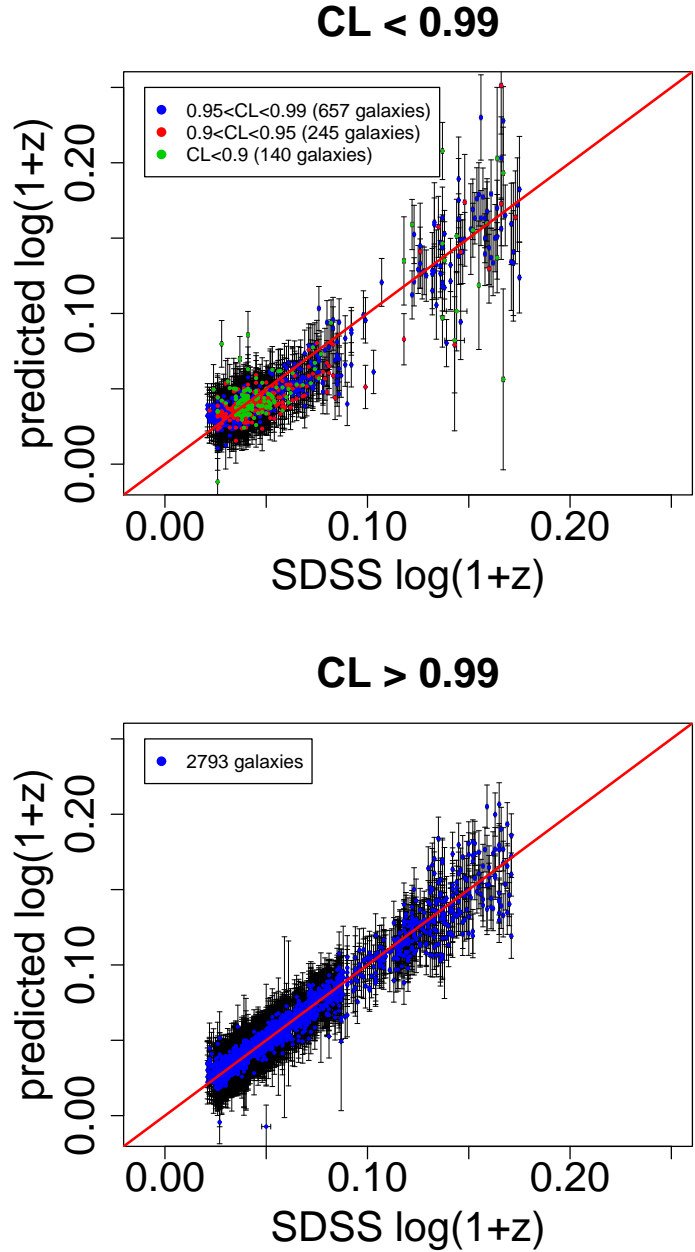}
\caption{
  Redshift predictions using diffusion map coordinates for galaxies
  with SDSS  CL $\le$ 0.99 (top)
  and CL $>$ 0.99 (bottom), each plotted against $z_{\rm SDSS}$.  
  Error bars
  represent 95\% prediction intervals.  Note that  CL $\le$ 0.99
  redshift predictions are based on the model trained on CL $>$ 0.99
  galaxies while CL $>$ 0.99 predictions are from 10-fold CV on CL
  $>$ 0.99 galaxies.  For most galaxies, our
  predictions are in close correspondence with SDSS estimates.}
\label{fig:zreg}
\end{figure}

\begin{figure}
\epsscale{0.6}
\plotone{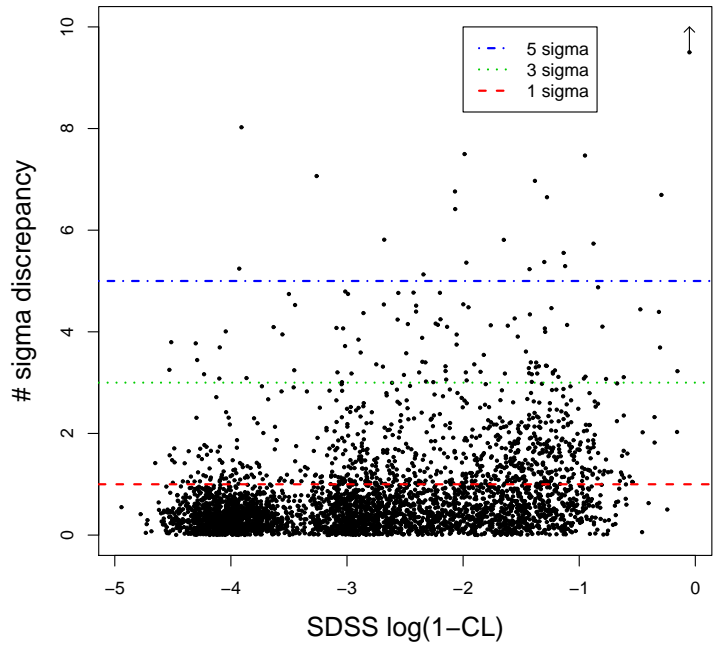}
\caption{Discrepancy between our predicted redshift values and $z_{\rm
    SDSS}$ estimates versus log(1-CL).  There is a 
  correlation of 0.392 between the amount of discrepancy and 1-CL, meaning
  that galaxies for which there are large differences between the two
  redshift estimates tend to be objects whose SDSS redshift
  confidences are low.  Horizontal lines denote 1, 3, and 5 $\sigma$
  disparities.  Small random perturbations have been added to duplicate
  log(1-CL) values to visualize galaxies with the same CL.  Galaxies with a
  CL of 1.00 are assigned mean log(1-CL) of -4.
}
\label{fig:cl}
\end{figure}

In Figure \ref{fig:zreg} we plot predictions and prediction
intervals for all galaxies in
our sample using our optimal diffusion map model.
(See Appendix \ref{sect:predint} for a discussion of prediction
intervals.)
Most of our predictions are in close correspondence with the SDSS
estimates.  We observe positive correlation in the amount of disparity between
our redshift estimates and SDSS estimates versus 1-CL (Figure
\ref{fig:cl}) meaning that galaxies for which our estimates disagree
with SDSS estimates are more likely to be galaxies with low CL.

\begin{figure}
\epsscale{0.6}
\plotone{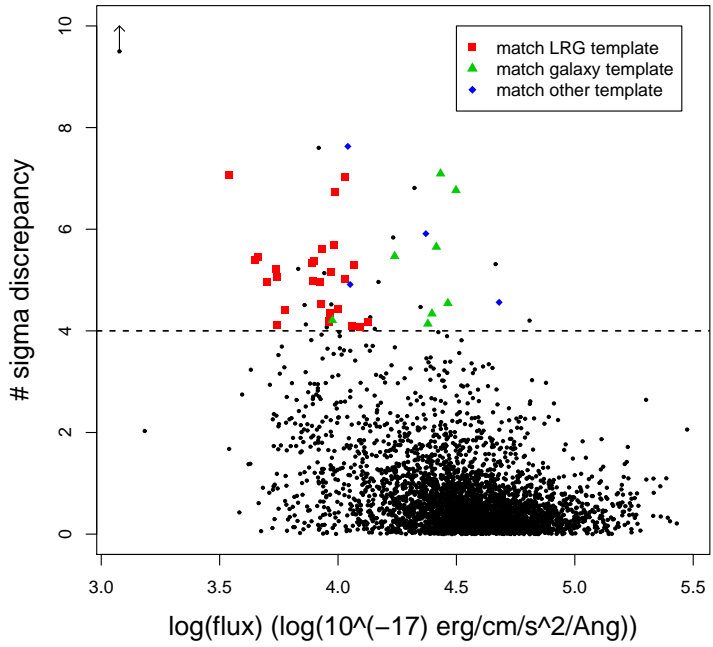}
\caption{Discrepancy between our predicted redshift values and $z_{\rm
    SDSS}$ versus log(flux) of the original spectra. There is a
  correlation of -0.327 between the amount of discrepancy and galaxy
  brightness. Galaxies can be detected as outliers even
    if they match well to their SDSS template (in color).  Low S/N
    can cause normal galaxies with correct SDSS redshifts to be labeled
    as outliers.  We also detect several
    physically interesting objects as outliers (see Figure \ref{fig:outliers}).
}
\label{fig:flux}
\end{figure}

There are 54 outliers at the $4\sigma$ level. Visual inspection of
their spectra indicates that 39 appear to fit the template assigned by
SDSS.  Of these, 27 are well-described by the LRG template.  In
Figure \ref{fig:flux} we show that most of the outliers that are
well-fit by their SDSS templates are faint objects.  A plausible
explanation for their classification as outliers is low S/N in their
measured spectra.  Faint galaxies with strong emission lines will
generally have accurate SDSS redshifts but can be outliers in
the diffusion map because noisy spectra induce higher Euclidean
distances.  In a future paper we will introduce a method to account
for errors in the original measured data
that corrects both for errors in Euclidean distance
computations and random errors in the diffusion map coordinates.

\begin{figure}
\epsscale{1}
\plotone{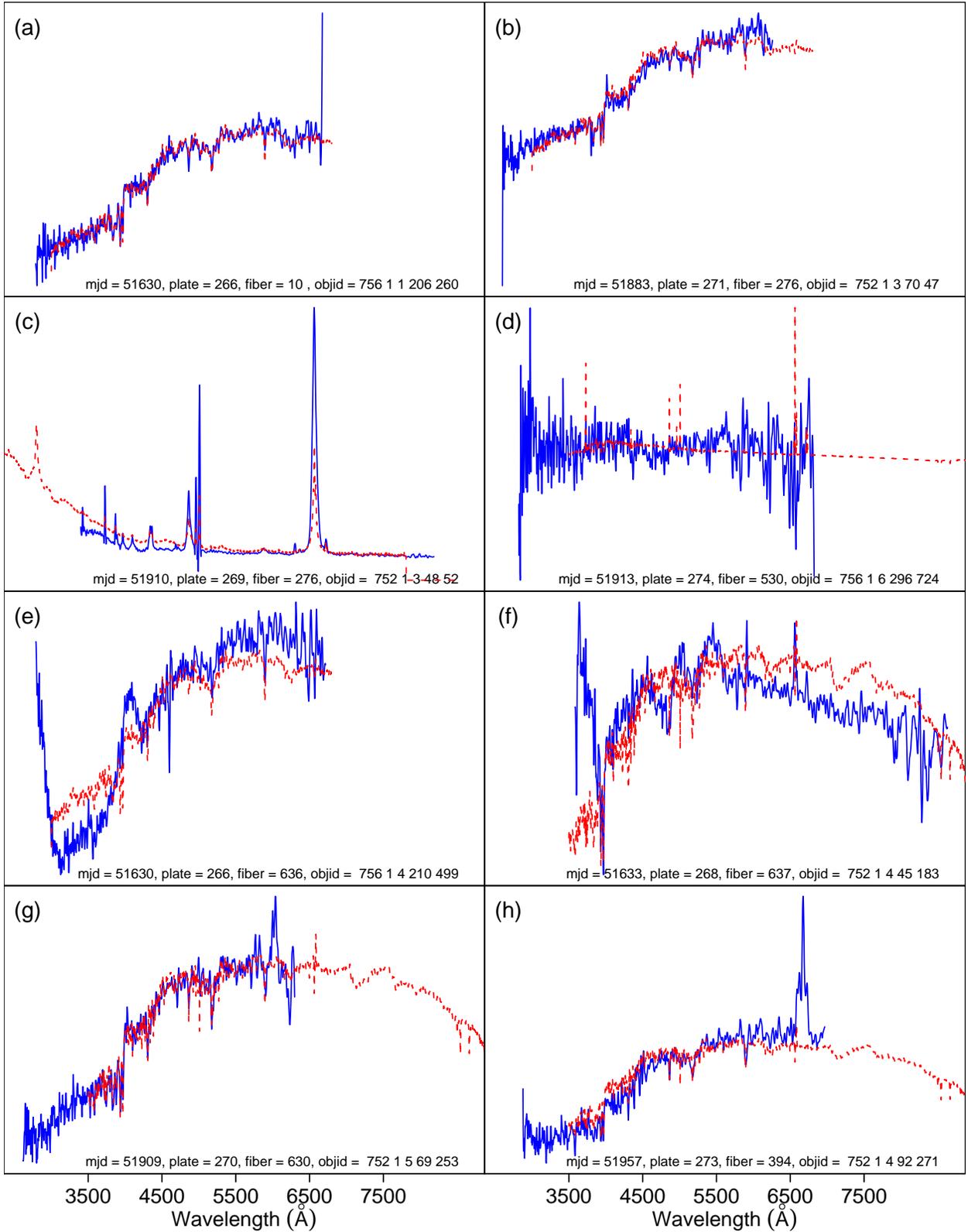}
\caption{Eight selected outliers with anomalous features.  Each
  spectrum (solid blue) is plotted along with its SDSS template match
  (dashed red).  Spectra are scaled to have the same sum of squared
  (smoothed) fluxes over the same range of wavelengths.  For a
  thorough discussion of
  these outliers see {\S}\ref{sect:anal}}
\label{fig:outliers}
\end{figure}

The 15 other outliers show interesting and/or anomalous features.
Four spectra appear to be LRG type galaxies with abnormal emission
and/or absorption features, of which at least two are likely
attributed to calibration errors (see Figure \ref{fig:outliers}a,b).
One spectrum is clearly a QSO (Figure \ref{fig:outliers}c), one shows
only sky subtraction residuals (Figure \ref{fig:outliers}d), and two others are
obvious mismatches to their SDSS
templates due to absorption lines whose depths do not match their
assigned template.  Four outliers have abnormal bumps (possible
continuum jumps due to instrumental artifacts, see Figure
\ref{fig:outliers}e,f) that appear like wide emission features.
One outlying galaxy has a spectrum that looks like
a late-type galaxy with no emission lines, meaning it is likely a
K+A post-starburst galaxy.  Another outlier has an anomalous emission
feature around 6000~\AA~ in rest frame (Figure \ref{fig:outliers}g).
This is a possible lens
galaxy, but was not selected by the Sloan Lens ACS Survey (SLACS;
\citeauthor{Bolton2006}) because
the feature in question
occurs in close proximity to strong sky lines at 8800~\AA~.  The final
outlier has a strong, wide emission feature in the
vicinity of H$\alpha$ but has no emission lines anywhere else in the
SDSS spectrum (Figure \ref{fig:outliers}h). 
None of the outlying spectra show conclusive evidence of a wrong SDSS redshift
measurement (except for the afore-mentioned sky spectrum, which we
detect as a 30 $\sigma$ outlier).


\subsection{Comparison With Other Methods}

As discussed in {\S}1, many authors have applied PCA to galaxy spectra
in an attempt to reduce the dimensionality of the data space, but few attempt
to find simple relationships between the reduced data and the physical
parameters of interest; these exceptions include 
\citeauthor{Li2005}, \citeauthor{Zhang2006}, and
\citeauthor{ReFiorentin2007}
In all three cases, the authors use
PCA to estimate stellar and/or galactic parameters that are traditionally
estimated by laboriously measuring equivalent widths and fluxes
of individual lines, just as we have used diffusion map eigenfunctions
to estimate redshift, a physical parameter usually estimated through
computationally intensive cross-correlation methods.
We stress three advantages of our approach over those employed by the
above authors: 
1) We achieve much lower prediction
error using diffusion map coordinates as compared to PCA,
2) we have an objective way of selecting the parameters of
 the model, and 3) we use a theoretically well-motivated regression
 model which takes statistical variations of the data into account and
 which unifies the data parameterization and regression algorithms.

The aim of \citeauthor{Li2005}~is to estimate, e.g., the velocity 
dispersion and reddening of a set of approximately 1500 galaxies
observed by SDSS.
They use PCA in two successive applications.
They first apply PCA
to the STELIB library to reduce 204 stellar spectra to 24 stellar eigenspectra.
These in turn are fit to SDSS DR1 spectra to create a library of 1016 
galactic spectra, which are reduced to nine galactic eigenspectra.
The authors then regress observed equivalent widths (EW) and fluxes of
H$\alpha$ upon these nine eigenspectra.
They determine the number of eigenspectra to retain 
by estimating noise variance in the stellar case
and by using the $F$ test to compute the significance of each additional
eigenspectrum in spectral reconstruction in the galactic case.  The latter
criterion however is not well-suited to the task of parameter
estimation because
the appropriate number of components in the regression model depends
on the complexity of the dependence of those parameters as a function
of the basis elements, not on the complexity of the original spectra.
For example, the dependence of the EW of H$\alpha$ on the PC basis
functions may be a simple, smooth function while the flux dependence
may be complex, bumpy relationship.  In this case, the optimal
regression model to predict EW would require fewer basis functions
than the optimal model for H$\alpha$ flux prediction.  Minimizing CV
risk would lead us to choose the correct number of basis functions for
each task, while the method of Li et al. would force us to use the same
(inappropriate) size for each model.

\citeauthor{Zhang2006} attempt to predict stellar parameters by
regressing on PC coefficients using a kernel regression model with a
 variable window width. In their paper, they do not specify how to
 select the window 
width (they introduce an arbitrary parameter $\lambda$) or how to
choose the correct number of PC basis functions (they use 3).
Their choice of a small
model size is likely due to the computational and statistical
difficulties that characterize kernel regression in high dimensions
\citep{Wasserman2006}.

\citeauthor{ReFiorentin2007}~attempt to estimate
stellar atmospheric parameters (effective temperature, surface gravity,
and metallicity) from SDSS/SEGUE spectra.
They use PCA for dimension reduction, but set $m$ to an
arbitrary value (e.g., 50). 
They then use an iterative, non-linear regression model (utilizing the
hyperbolic tangent function; see \citealt{Bailer-Jones2000}),
with an error function based on the residual sum-of-squares plus
a regularization term (see their equation 2). Again, the
choice of the regularization parameter is not justified.
We find that when applied to the same data
set of galaxy spectra, their model does not achieve lower CV risk than
our model for different choices of regularization parameter and model size.

\section{Summary}

\label{sect:summary}

The purpose of this paper is two-fold.
First, we introduce the diffusion map method for data parametrization
and dimensionality reduction. We show
that for the types of high-dimensional and complex data sets
often analyzed in the astronomy, diffusion map can yield
far superior results than commonly-used methods such as PCA.  Moreover,
the simple, intuitive formulation of diffusion map as a method that
preserves the local interactions of a high-dimensional data set makes the
technique easily accessible to scientists that are not well-versed in
statistics or machine learning.

Second, we present a fast and powerful eigenmode-based framework for
estimating physical parameters in databases of high-dimensional
astronomical data.  In most astrophysical applications, PCA is used as
a data-explorative tool for dimensionality reduction,
with no formal methods
and statistical criteria for regression, risk estimation and selection
of relevant eigenvectors. Here we propose a statistically rigorous,
unified framework for
regression and data parameterization.  Our proposed regression model
combines basis functions in a simple and statistically-motivated
manner while our clear objective of risk minimization drives the
estimation of the model parameters.  Again, the simplicity of the
proposed method will make it appealing to the non-specialist.

 We apply the proposed methodology to predict redshift for a sample of
 SDSS galaxy spectra, comparing the use of the proposed regression
model with PCA basis functions versus diffusion map basis functions. 
We find that the prediction error for the diffusion-map-based approach
is markedly smaller than that of a 
similar framework based on PCA. Our techniques are also more robust
than commonly used template matching
methods because they consider the structure of the entire
high-dimensional data set when reparametrizing the data.
Statistical inferences are based on this learned structure,
instead of considering each data point separately in an object-by-object
matching algorithm as is currently used by SDSS and commonly employed
throughout the astronomy literature.
Work in progress extends our approach to
photometric redshift estimation and to the estimation of the
intrinsic parameters (e.g., mean metallicities and ages) of galaxies.

\begin{acknowledgments}
The authors would like to thank Jeff Newman for helpful conversations.
This work was supported by NSF grant \#0707059 and ONR grant N00014-08-1-0673.
\end{acknowledgments}

\appendix

\section{Principal Components Analysis}
\label{sect:pca}

We first center our data (the normalized spectra with $p$ wavelength bins) so that $\frac{1}{n} \sum_{i=1}^{n} {\bf x}_i = 0$. The centered observations ${\bf x}_1, {\bf x}_2, \ldots {\bf x}_n \in \mathbb{R}^p$ are then stacked into the rows of an $n \times p$ matrix ${\bf X}$. Note that the sample covariance matrix of $\bf x$ is given by the $p \times p$ matrix ${\bf S}= \frac{1}{n}{\bf X}^T{\bf X}$. In Principal Component Analysis (PCA), one computes the eigenvectors of the covariance matrix that correspond to the $m < p$ largest eigenvalues; denote these vectors by ${\bf v}_1, \ldots, {\bf v}_m \in \mathbb{R}^p$. In a PC map, the projections of the data onto these vectors are then used as new coordinates; i.e. the PC embedding of data point ${\bf x}_i$ is given by the map
$$ {\bf x}_i \mapsto \Psi_{\rm PCA}({\bf x}_i)=({\bf x}_i \cdot {\bf v}_1, \ldots, {\bf x}_i \cdot {\bf v}_m).$$ 
These projections are sometimes referred to as the principal components of ${\bf X}$.

Algorithmically, the PC embedding is easy to compute using a singular value decomposition (SVD) of ${\bf X}$:
$$ {\bf X=U D V}^T. $$
Here ${\bf U}$ is an $n \times p$ orthogonal matrix,  ${\bf V}$ is a $p \times p$ orthogonal matrix (where the columns are eigenvectors ${\bf v}_1, \ldots, {\bf v}_p$ of ${\bf S}$), and ${\bf D}$ is a $p \times p$ diagonal matrix with diagonal elements $d_1 \geq d_2 \ldots \geq d_p \geq 0$ known as the singular values of ${\bf X}$. Since ${\bf XV}={\bf UD}$, the PC embedding of the $i$:th data point in $m$ dimensions is given by the first $m$ elements of the $i$:th row of ${\bf UD}$.

\section{Prediction Intervals for Spectroscopic Redshift Estimates}

\label{sect:predint}

In any one fold of a ten-fold regression analysis, we fit to 90\% of the data,
generating predictions and prediction intervals 
for the 10\% of the data withheld from the analysis.  A prediction interval
is {\it not} a confidence interval; the former 
denotes a plausible range of values for a single observation, whereas the
latter denotes a plausible range of values for a parameter of the
probability distribution function from which that single observation is
sampled (e.g., the mean).

Let $\bf X$ and $\bf \tilde X$ represent the matrices of independent variables 
included in, and withheld from, regression analysis, respectively.  For
instance,
\begin{eqnarray}
{\bf \tilde X}~=~
\left(
\begin{array}{cccc}
\psi_1(x_1) & \cdots & \cdots & \psi_m(x_1) \\
\vdots      & \vdots & \vdots & \vdots \\
\psi_1(x_n) & \cdots & \cdots & \psi_m(x_n)
\end{array}
\right) \,, \nonumber
\end{eqnarray}
where $n$ is the number of withheld data and $m$ the number of
assumed basis functions.  (Here, we leave out factors of
$\lambda_j^t$, which are subsumed into the estimated
regression coefficients ${\widehat \beta}_j$.)  The vector of 
redshift predictions for the withheld data is thus
\begin{eqnarray}
{\widehat z}~=~{\bf \tilde X} {\widehat \beta} \,, \nonumber
\end{eqnarray}
where $\widehat \beta$ is estimated from ${\bf X}$
while the vector of half-prediction intervals is given by
\begin{eqnarray}
t_{\alpha/2,N-n-2} \widehat{\sigma} \sqrt{ {\bf \tilde X} \left( {\bf X}^T {\bf X} \right)^{-1} {\bf \tilde X}^T + 1 + \frac{1}{N-n} } \,,
\label{eqn:predint}
\end{eqnarray}
where $\widehat{\sigma}$ is the estimated standard deviation of the
random noise $\epsilon$ in the relationship $Y = r({\bf X}) + \epsilon$,
estimated from the residuals of the regression of $Y$ upon ${\bf X}$,
$t_{\alpha/2,N-n-2}$ is the critical t-value for a two-sided
100(1-$\alpha$)\% prediction interval,
and $N$ is the total number of data points.  Equation (\ref{eqn:predint}) is 
a multi-dimensional generalization of, e.g., equation (2.26) of 
\citet{Weisberg2005}, taking into account that the mean of $\psi({\bf x})$ is
zero.


\clearpage

\clearpage


\clearpage

\clearpage

\clearpage

\clearpage

\clearpage

\clearpage

\input{tab1}

\end{document}

%% file: tab1.tex
\begin{deluxetable}{lcccccc}
\tablenum{1}
\tablecaption{Parameters of Optimal Regression on $\log_{10}(1+z_{\rm SDSS})$}
\tablewidth{0pt}
\tablehead{
 & & & & \multicolumn{3}{l}{Number of Outliers} \\ & \colhead{$\epsilon_{opt}$} & \colhead{$m_{opt}$} & \colhead{$\widehat{R}_{CV}(\epsilon_{opt},m_{opt})$\tablenotemark{a}} & \colhead{$3\sigma$} & \colhead{$4\sigma$} & \colhead{$5\sigma$}
}
\startdata
Diffusion Map & .0005 & 127 & 0.1341 & 115 & 54 & 20 \\
PC & -- & 88 & 0.1931 & 109 & 55 & 20
\enddata\tablenotetext{a}{Prediction risk estimated via cross-validation; see equation (\ref{eqn:orthoreg}) and subsequent discussion.}
\label{tab:zreg}
\end{deluxetable}